\documentclass[preprint, single-spaced,showkeys,showpacs]{revtex4}
\usepackage{graphicx,amsmath,bbm,bm,array,subfigure} 
\usepackage{appendix}
\usepackage{lipsum}
\usepackage[linktocpage=true,colorlinks=true,linkcolor=blue,citecolor=blue,urlcolor=blue]{hyperref}
\usepackage{MnSymbol}
\def\nn{\nonumber\\}

\usepackage{graphicx,amsmath,bbm,bm,array,subfigure} 
\usepackage{appendix}
\usepackage{lipsum}
\usepackage[linktocpage=true,colorlinks=true,linkcolor=blue,citecolor=blue,urlcolor=blue]{hyperref}
\usepackage{MnSymbol}
\def\nn{\nonumber\\}

\begin{document}
\title{Airborne virus transmission under different weather conditions}
\author{Santosh K. Das}
\email{santosh@iitgoa.ac.in}
\affiliation{School of Physical Sciences, Indian Institute of Technology Goa, Ponda-403401, Goa, India}
\author{Jan-e Alam}
\email{jane@vecc.gov.in}
\affiliation{Variable Energy Cyclotron Centre, 1/AF Bidhan Nagar, Kolkata- 700064, India}
\affiliation{Homi Bhabha National Institute, Training School Complex, Mumbai - 400085, India}
\author{Salvatore Plumari}
\email{salvatore.plumari@dfa.unict.it}
\affiliation{Department of Physics and Astronomy, University of Catania, 
Via S. Sofia 64, I-95125 Catania, Italy}
\affiliation{Laboratori Nazionali del Sud, INFN-LNS, Via S. Sofia 62, I-95123 Catania, Italy}
\author{Vincenzo Greco}
\email{greco@lns.infn.it }
\affiliation{Department of Physics and Astronomy, University of Catania, 
Via S. Sofia 64, I-95125 Catania, Italy}
\affiliation{Laboratori Nazionali del Sud, INFN-LNS, Via S. Sofia 62, I-95123 Catania, Italy}
\def\zbf#1{{\bf {#1}}}
\def\bfm#1{\mbox{\boldmath $#1$}}
\def\hf{\frac{1}{2}}
\def\sl{\hspace{-0.15cm}/}
\def\omit#1{_{\!\rlap{$\scriptscriptstyle \backslash$}
{\scriptscriptstyle #1}}}
\def\vec#1{\mathchoice
        {\mbox{\boldmath $#1$}}
        {\mbox{\boldmath $#1$}}
        {\mbox{\boldmath $\scriptstyle #1$}}
        {\mbox{\boldmath $\scriptscriptstyle #1$}}
}
\def \be{\begin{equation}}
\def \ee{\end{equation}}
\def \beqa{\begin{eqnarray}}
\def \eeqa{\end{eqnarray}}
\def \nn{\nonumber}
\def \pd{\partial}

\def \la{\langle}
\def \ra{\rangle}

\begin{abstract}
The  COVID19 infection is known to disseminate through droplets ejected by infected 
individuals during coughing, sneezing, speaking and breathing. The spread of the infection 
and hence its  menace depend on how the virus-loaded droplets evolve in space and time
with changing environmental conditions. In view of this, we investigate the evolution of 
the droplets  within the purview of the Brownian motion of the evaporating droplets in the air 
with varying  weather conditions under the action of gravity. We track the movement of the 
droplets till either they gravitationally settle on the ground or evaporate to aerosols of 
size 2$\mu$m or  less. Droplets with radii $2 \mu$m or less  may continue to
diffuse and remain suspended in the air for long time. The effects of relative humidity and 
temperature on the evaporation are found to be significant. We note that 
under strong flowing conditions  droplets travel large distances. It is found that the bigger 
droplets fall on the ground due to the dominance of gravity over the diffusive force
despite the  loss of mass due to evaporation. The smaller evaporating droplets 
may not settle on the ground but remain suspended in the air due 
to the dominance of the diffusive force. The fate of the intermediate size droplets depends 
on the weather conditions and play  crucial roles in the spread of the infection. These 
environment dependent effects indicate that the maintenance of physical separation to 
evade the virus is not corroborated, making  the use of face mask indispensable. 
\end{abstract}
\maketitle
\section{Introduction}
It is well known that droplets released by infected 
persons through coughing, sneezing,
speaking or breathing contain microorganism (bacteria, virus, fungi, etc) causing
a large number of diseases~\cite{cflugge,Wells}. 
Several decades ago it was considered that the infections which are contained 
within the droplets are airborne~\cite{Gelfand}.  Recently ten scientific reasons 
have been provided in support of SARS-COV-2 as an airborne infection~\cite{10reasons}
(see also ~\cite{review} for review).
The transmission routes ~\cite{kutter} of the virus crucially  depends on how 
the droplets evolve in space with time under the action of three competitive forces - 
gravitational force which is opposed by the drag and diffusive forces. For example, big droplets
settle gravitationally on different surfaces like, hand rail, door handle, tables,
etc and eventually infect through direct contact.  The small droplets either directly
discharged or formed due to evaporation or fragmentation of the big droplets  
remain suspended in the air for long time due to the dominance of the diffusive force over 
the gravitational which infect healthy persons through inhalation. 
It still remains as a big challenge to 
understand the density of SARS-COV-2 virus in the air~\cite{mpan},
their capability to infect~\cite{bynumbers}, their survivability
on different surfaces~\cite{doremalen,AA} and the relative weight these factors. 
Therefore, these factors limit our ability to evaluate the risk~\cite{Nazhu}.
As the saliva of the infected persons contains the Coronavirus~\cite{kkwang,lazzi}
hence it can be transmitted through the process of 
respiration~\cite{richard,herfst}, speaking~\cite{kkwang,asadi,stadnytskyl},
coughing and sneezing. The increase in emission of pathogens with the
loudness in speaking which may depend on some unknown physiological factors
varying among individuals has also been reported ~\cite{asadi}.

Recently it has been reported that the droplets and aerosols
(droplets with size $<5 \mu$m~\cite{seta}) can travel a distance much larger 
than the prescribed 6 feet and remain suspended in the air for hours
which is supported by the fact that the  SARS-COV2 RNA is recovered in 
air sample~\cite{jama20}.
The propagation and aerosolisation of the droplets have been investigated within the purview of 
statistical mechanics and fluid dynamics ~\cite{dbouk,mittal,MRP}. The Euler-Lagrange equation 
has been applied to study the effects of the size, ejection velocity and angle of emission of 
the droplets on their trajectories. 
However, the application of the stochastic statistical mechanics~\cite{Reif,pathria} is particularly 
crucial to study the motion of small droplets and aerosols   
for which the airborne transmission
turns out to be very vital. The Brownian motion of the aerosols in the air-bath can be studied 
by solving Langevin stochastic differential equation~\cite{SKDas}. 
In the Euler-Lagrange approach the stochastic motion is not included.

Here we investigate the spread of these ejected  droplets  in the  neighborhood of the 
infected individual under varying weather conditions. The result will be useful 
in planning the preventive strategies at different climatic conditions. 
The droplets ejected  interact with the molecules of the still/flowing air
at temperature ($T$), relative humidity ($\textrm{RH}$)  
in the presence of gravity. The interacting forces between the droplets and the 
air molecules are changing continuously as the molecules are changing their coordinates 
continuously  with time. This  makes the problem very complex making it not 
solvable exactly. The spread of the droplets, however, will depend on these interactions.
Under such circumstances the air molecules can be regarded as forming a thermal bath 
where the droplets are executing Brownian motion  with its changing mass due to evaporation.
The interaction of the evaporating droplets with the bath can then be 
grouped into drag and diffusive forces quantified through the drag and diffusion
coefficients. Therefore, there are three
types of force acting on a droplet which are: (i) drag and (ii) diffusive  forces
between the droplet and the air molecules and the (iii) Newtonian gravitational force on
the droplets due to their non-zero mass. However, it is crucial to note that droplets
are undergoing loss of mass due to evaporation and hence 
the the gravitational force is changing with time too. 

In the present work we investigate the propagation of the  virus-containing droplets 
subject to continuous evaporation 
by solving the Langevin stochastic differential equation of statistical 
mechanics~\cite{Reif,pathria}
coupled with the equation that governs the evaporation of the droplets.  
The Langevin equation is applicable in the present context as the mass of the droplets are 
much higher than the mass of oxygen and nitrogen molecules forming the bulk of the air.  

It is crucial to mention here that the flow of the 
droplet in the air may either be laminar or turbulent
depending on the situation.  In the present work
this fact has been taken into consideration with appropriate
parametrization Reynold's number ($Re$) which appear through  
the drag force~\cite{Holterman}. 
The inclusion of all the forces mentioned above enable us to study the trajectories of droplets with
a wide range of sizes. It will be interesting to study how the evaporating droplets
evolve in space and time under the influence of gravitation which
will act to pull the droplet on the ground in contrast to the diffusive and drag forces
which will prevent it to fall on the ground. The big (hence massive) droplets is expected to
settle gravitationally quickly and the smaller one is expected to remain suspended
in the air for longer time. However, under evaporation the droplets  suffer
continuous loss of mass and consequently a droplet which will otherwise fall 
on the ground due gravitation may not do so but remain suspended as smaller droplets/aerosol
or isolated virus for longer time before complete decomposition.  

The present work is a sequel of previous publication~\cite{SKDas}. In contrast 
to the earlier work, here we include the process of evaporation of the droplets and its flow beyond the 
laminar  region.  
The paper is organized as follows. In the next section the solution of the Langevin equation
coupled with the equation that governs the evaporation has been presented. 
Section III contains the results and section IV is devoted to summary and discussions.

\section{Solving the Langevin equation in the presence of evaporation}
The Langevin equation,  governing the motion of the droplet of mass ($M$) 
in the still air in the presence of gravitational field~\cite{Reif} is given by: 
\begin{equation}
\frac{dr_i}{dt} = v_i 
\label{eq1}
\end{equation}
\begin{equation}
M \frac{dv_i}{dt} =  -\lambda v_i + \xi(t) +F^G 
\label{eq2}
\end{equation}
In Eqs.~\ref{eq1} and \ref{eq2} the $dr_i$ and $dv_i$ are the shifts of the coordinate and velocity in 
each discrete time step $dt$, $i(=x,y,z)$ stands for the Cartesian components of the 
position and velocity vectors. The 
$\lambda$ in Eq.~\ref{eq2} is the drag coefficient which will be fixed soon. 
The first term in the right hand side of Eq.~\ref{eq2} represents 
the dissipative force and the second term 
stands for the diffusive (stochastic) force.
$\xi(t)$  is  also  called  noise due  to  its  stochastic  nature.
We study the evolution with a white noise ansatz for $\xi(t)$, {\it i.e}  
$\la{\bf \xi}(t)\ra = 0$ and
$\la {\bf \xi}(t) {\bf \xi}(t')\ra = \kappa\delta(t-t')$,
where $\kappa$ is the diffusion coefficient which regulates the $\xi(t)$.
White noise describes a fluctuating field without memory, whose correlations have
an instantaneous decay called  $\delta$ correlation. 
The third term in Eq.~\ref{eq2}, $F^G$ represents the gravitational
force $(=Mg$, $g=9.8$ m/s$^2$) acting on a droplet of mass $M$ 
which changes with time due to the evaporation.
The mass of the droplet is related to its diameter ($D$)  as
$M=\pi D^3\rho_L/6$ ion where
$\rho_L$ is the density of the evaporating liquid.

The rate of decrease of the diameter $D$ of a spherical liquid 
drop due to evaporation is given by
~\cite{Holterman,Kukkonen}:
\be
\frac{dD}{dt×}=-\frac{4M_LD_v}{D\rho_L R T_f×}\Delta p(1+0.276Re^{1/2}Sc^{1/3})
\label{eq3}
\ee
In Eq.~\ref{eq3}, $M_L$ ($=0.018$ kg/mol) is the molecular weight 
of the liquid, $D_v$ is the diffusion coefficient 
of the vapor molecules in the saturated film around the droplets 
and $T_f$ is the average temperature 
of the film formed around the droplets due to evaporation,  
$R=8.3144$ J/(mol K) is the gas constant and 
$Re$ is the Reynold's number~\cite{Landau}, given by:
\be
Re=\frac{D\rho_a\,v}{\eta_a×}
\ee
where $v$ is the relative velocity of the droplet with respect to the surrounding air, $\rho_a$ is the 
density and $\eta_a$ is the viscosity of the air at temperature $T_f$.
$Sc$ is the Schmidt's number, given by:
\be
Sc=\frac{\eta_a}{\rho_a D_v×}
\ee
$\Delta p$ is the difference between the vapor pressure near the 
droplet and in the atmosphere which acts as 
the driving force for the transport of vapor away from the droplet surface.   
$\Delta p$ can be related to the saturated vapor pressure at ambient temperature $(T)$ 
and wet-bulb temperature ($T_w$) as:
\be
\Delta p=p_{sat}-p=\gamma (T-T_w)
\ee
where $p_{sat}$ is the vapor pressure near the surface of the droplet,  
$p$ is the vapor pressure in the ambient air and $\gamma$ is approximately  constant ($\sim$ 67 pa/K).
$T_w$ can be  expressed in terms of $T$ and $\textrm{RH}$ as:
\be
T_w=T-[(a_0+a_1T)+(b_0+b_1T){\textrm{RH}}+(c_0+c_1T){\textrm{RH}}^2]
\ee
with $a_0$=5.1055, $a_1$=0.4295, $b_0$=-0.04703, $b_1$=-0.005951, 
$c_0$=-0.00004005 and $c_1$=0.0000166,
for further details we refer to ~\cite{Holterman}.

The trajectories of the droplets depend on the ejected 
velocity (initial) and on the flow velocity of the ambience. 
It is obvious that droplets in still and flowing air conditions 
[such as in a air conditioned (AC) room or open air] will
follow different paths.
The results for both the scenarios of still and flowing air
are provided below.  
The air flow velocity has been taken into account  
through  the Galilean transformation of the Langevin equation.
Since the virus carrying droplets follow different trajectories
in still and flowing air the preventive strategies for
indoor and out door conditions should take care of this fact.
Here we consider a velocity profile for the air flow  
as: $u(x)=u_0(1-\frac{x}{x_{\text{max}}})$
with its upward and downward components as zero to serve this purpose,
where  $x$ is the running coordinate,
$u_0$ is the peak value of $u(x)$ at $x=0$ which will be   
varied to check the sensitivity of the results.
Here the $x_{\text{max}}$ is the maximum value of $x$, which may be 
restricted to the size of an AC room in indoor condition. 
We will also consider constant (no dependence on spatial coordinates)
air flow velocity to calculate the distance followed by ejected droplets 
in outdoor condition.  

The value of the drag coefficients, $\lambda$ is estimated by using 
the relation, 
\begin{equation}
\lambda=\frac{1}{2}C_D\rho_a\,S\,v
\label{eq4}
\end{equation}
where $S$  is the projected 
cross sectional area of the spherical droplet of diameter $D$.
The following expression is used for $C_D$~\cite{hongping} to extend the validity of the model
beyond the region of laminar flow.
\begin{equation}
C_D=\frac{24}{R_e}+\frac{6}{1+\sqrt{R_e}}+0.4
\end{equation}
The diffusion coefficient is obtained by using the Einstein relation
\cite{pathria}, $\kappa=k_BT\lambda$,
where $k_B=1.38\times 10^{-23} J/^{\circ}K$, is the Boltzmann constant.
It may be noted that for $C_D=24/Re$, the well known Stokes law is recovered.

The initial spatial coordinate of the droplet is , $x=y=0$ and $z=H_0$, where $H_0$ is the height
(1.7 meter) at which the droplet is released (nose/mouth), {\it i.e.}
$(x,y,z)=(0,0,1.7{\text {meter}})$, is the point of ejection.  
The eqs.~\ref{eq1}, \ref{eq2} and \ref{eq3}  
have been simultaneously solved  numerically
to obtain the trajectories of the droplets  with the inputs mentioned above.
We use the Monte-Carlo method to sample the initial phase-space distribution of the droplets 
\cite{SKDas,mc1,mc2}. 
The calculation is done for different 
initial radii of droplets varying  from 10 $\mu$m to 200$\mu$m~\cite{RR} and
the ejection velocity, $V_0=21$ m/s~\cite{VE}. 
The initial velocity (at $t=0$) is uniformly distributed in the 
$x-y$ plane with $v_z=0$. 
We use the Euler method at second order 
including the diffusive (stochastic) force. We have checked the convergence and stability of 
our solutions with respect to the time step using analytical solutions of the Langevin equation 
for simple configurations and then using the conditions relevant for the present
work to get the trajectories of the droplets.  
This numerical technique has been used to reproduce available results in the literature 
(for example, the results displayed Fig.11(a) of Ref.~\cite{hongping})
under the same conditions.

The solutions of  Eqs.~\ref{eq1}, \ref{eq2} enable us to calculate
the horizontal distance ($L(t)$), traveled by the droplets from the point of ejection
as a function of time as: $L(t)=\sqrt{x(t)^2+y(t)^2}$.
Its maximum value of $L(=L_{\text{max}})$ dictates the stationary
distance that to be maintained between infected and healthy persons to prevent 
the virus.  The solution of these equations can also be used to estimate
the maximum time ($t_{\text{max}}$) of suspension of the droplets. In the following
section, results for both $L_{\text{max}}$ and $t_{\text{max}}$ have been presented. 

\section{Results}
The virus carrying droplets are ejected with different sizes
and initial velocity in the ambience in  widely varying 
climatic conditions. Therefore, results obtained 
by solving Eqs.~\ref{eq1}, \ref{eq2} and ~\ref{eq3}
with different 
size of the droplets under different meteorological 
conditions are exhibited here to understand the prevention measures 
to be adopted to avoid the infection. We assume that the ejection velocity 
is 21 m/s (unless stated otherwise)
which is close to its highest possible value found experimentally
in Ref.~\cite{VE,zhu}. For given weather condition, droplet with highest ejection velocity  
will travel maximum distance. Consequently, it will also decide the maximum
social distance to be maintained to avoid the infection.
The weather effects have been taken into consideration through temperature, 
relative humidity and wind flow. The effects of evaporation and correction
to drag force for flow beyond the laminar region have been included as
mentioned above.   
The value of temperature is $20^\circ$ wherever it is not 
stated explicitly. 

\begin{figure}[ht]
\includegraphics[width=18pc,clip=true]{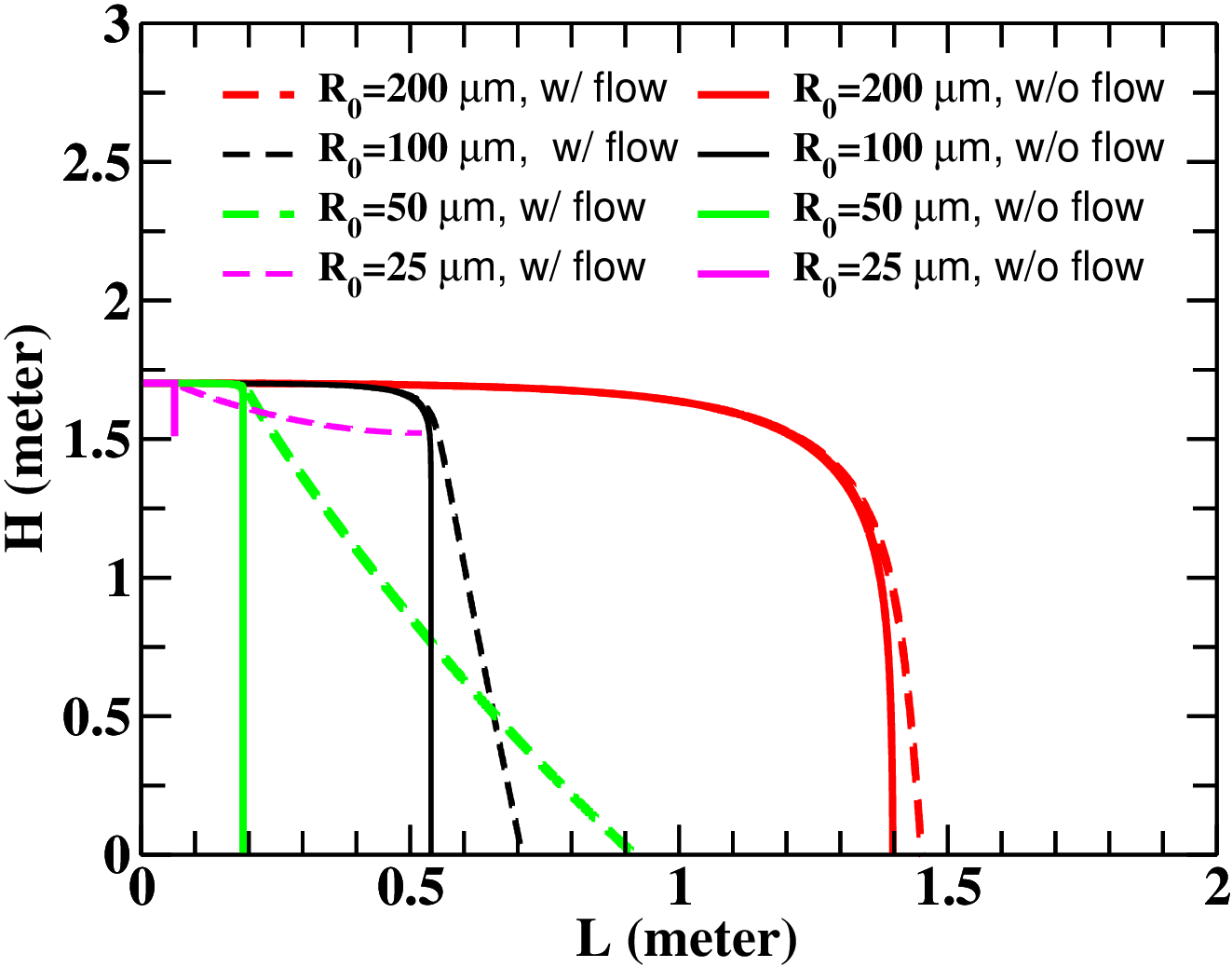}\hspace{2pc}
\caption{The variation of the  height ($H(t)$) with the horizontal
distance ($L(t)=\sqrt{x^2+y^2}$) is shown for droplets
of various radii  for  the initial ejection velocity, $V_0=
21$ m/s and the peak value of the wind flow velocity, $u_0$=0.1 m/s.
The results contain the effects of evaporation at ambience temperature $20^\circ$ C
and $\textrm{RH}=60\%$.
}
\label{fg1}
\end{figure}

In Fig.~\ref{fg1} the variation  of the height ($H$) with longitudinal 
distance ($L$) for  droplets of different radii released at a height of 1.7 meter  
is depicted. It is seen that a large droplets of radius 200 $\mu$m 
propagates upto a distance of
1.4 meter horizontally  in still air condition and 1.45 meter in wind flow condition 
with the peak velocity of the wind, $u_0=0.1$ m/s.  
The interplay between the 
ejection velocity and the wind velocity can be understood from the following
discussions. It may be mentioned here that the terminal or sedimentation velocity ($v_t$) 
of a droplet of diameter $D$ in air, obtained by balancing the drag plus the buoyant forces with
the gravitational force, is given by
\begin{equation}
v_t=\sqrt{\frac{4}{3}\,g\frac{\rho_L-\rho_a}{\rho_a}\frac{D}{C_D}}
\end{equation}
which reduces to the well known expression for $v_t$ in the 
laminar flow region 
(with $C_D=24/R_e$) as: 
\begin{equation}
v_t=\sqrt{\frac{g}{18}\,R_e\frac{\rho_L-\rho_a}{\rho_a}}
\end{equation}

The effects of wind flow becomes insignificant if the 
terminal velocity of a droplet is larger than the wind flow velocity. 
Here the wind velocity acts along  the horizontal direction whereas
the sedimentation or terminal velocity acts along the downward vertical direction.
If the terminal/sedimentation velocity dominates then the resultant will be tilted toward
the downward direction which will force the droplet to settle on the ground.
For a large droplet of size 
200 $\mu$m the $v_t$($\sim \sqrt{D}$) is larger than $u_0(=0.1)$ which
makes the effect of wind flow marginal as observed in the results
displayed in Fig~\ref{fg1}. 
The mass dependence of the distance traveled by a droplet 
can be understood simply by ignoring the diffusive 
force and assuming the motion of
the droplet in $x-z$ plane. In such situation the trajectory of
the droplet in still air is given by,
\begin{equation}
z=z_0+g\frac{M^2}{\lambda^2}ln(1-\frac{\lambda x}{Mv_x^0})+x\frac{\lambda v_z^0+Mg}{\lambda v_x^0}
\label{trajectory}
\end{equation}
where $z_0$ is the initial value of $z$ and $v_z^0$ ($v_x^0$) is the initial 
value $z$  ($x$) component of velocity.  
The maximum allowed value of $x$, $x^{\text max}$ is determined by this solution as,
$x^{\text max}=M v_x^0/\lambda$. It is important to note here that 
$\lambda \sim R \sim M^{1/3}$ for fixed density. Therefore, 
$x^{\text max}\sim M^{2/3}$, indicates that heavier droplets 
travel more distance than the lighter droplets for given $v_x^0$. This is 
supported by our numerical calculations (see also~\cite{zohdi}). 
The trajectory is terminated at $x^{\text max}$ because at this point the
forward momentum is completely spent to overcome the drag (frictional) 
force exerted by the air.
A droplet of size 100 $\mu$m travels a distance 0.55 meter and 0.7 meter respectively
in still and wind flow conditions implying that the wind flow
has larger effects on smaller droplet as their terminal velocity is smaller.  
The effects of wind flow is weaker for 50$\mu m$ droplet.
We observe that the droplets of size 25 $\mu$m evaporates 
to aerosols before settling on the ground. The  
diffusive force dominates over the gravitational force 
for small droplets which enable them to
float in the air for longer time (see also~\cite{SChatterjee}). 
For higher wind speed the effects of terminal velocity
is small, for example, a droplet of radius $100 \mu$m can travel 
about 7.4 meter for constant wind speed of 2 m/s as seen below (also see Ref.~\cite{HLi}). 
\begin{figure}[ht]
\includegraphics[width=7.0cm]{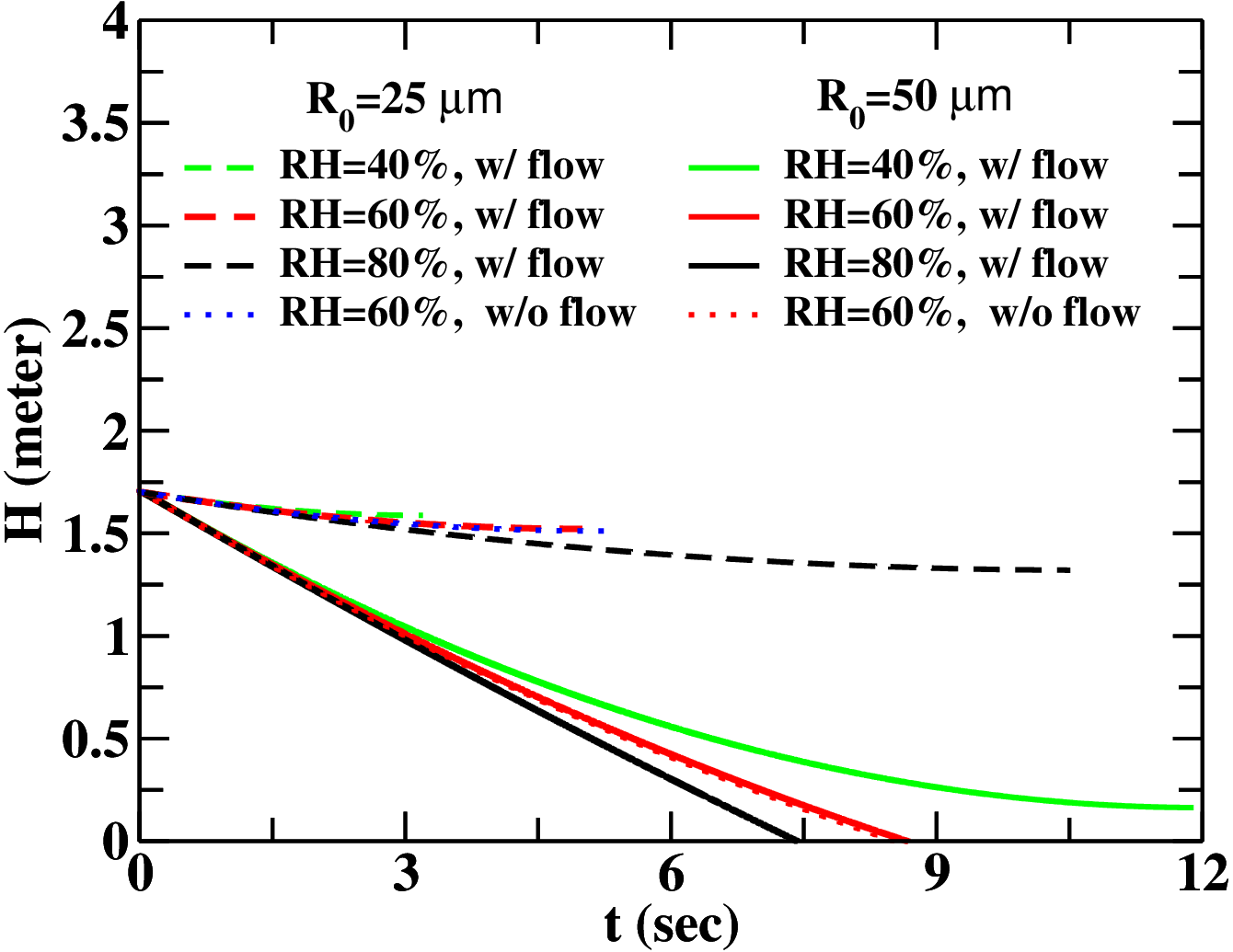}
\includegraphics[width=7.0cm]{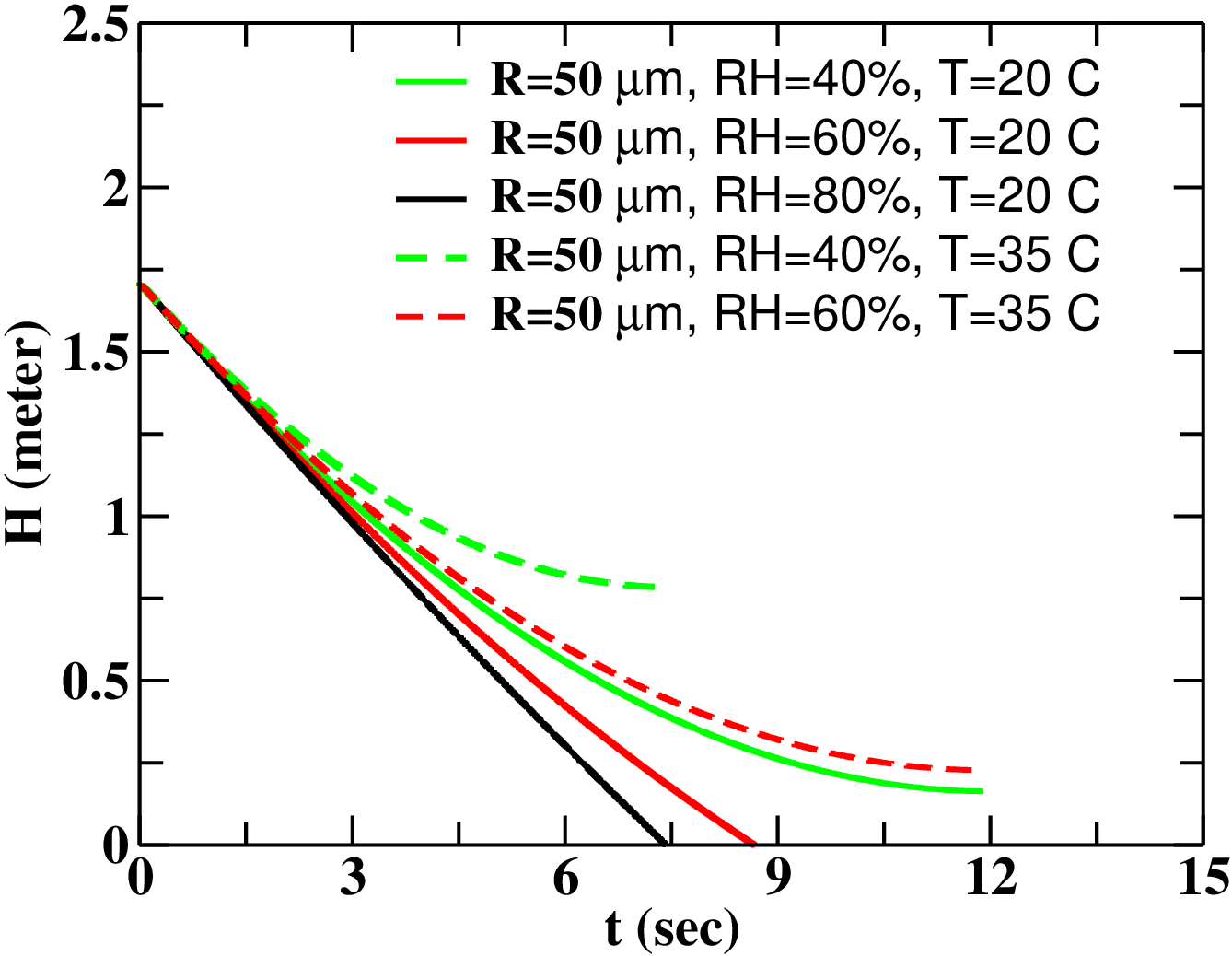}
\caption{(Color online) 
Left panel:The change in the height, $H(t)$, as a function of time 
for droplets of different sizes have been depicted. Here the
initial ejection velocity, $V_0=21$ m/s and the 
peak value of the wind flow velocity, $u_0$=0.1 m/s at $T=20^\circ$ C.
Right panel: Same as left panel showing the sensitivity of results
on the ambient temperature ($T=20^\circ$ C and $35^\circ$ C.
The results are derived with the inclusion of evaporation.}
\label{fg2}
\end{figure}

Fig.~\ref{fg2} illustrates the time that droplets of different sizes take to
settle gravitationally under different conditions of relative humidity and wind flow. 
We find that the droplets at smaller $\textrm{RH}$  evaporate  and  do not
settle on the ground due to weaker effect of  gravity. 
For larger value of $\textrm{RH}$ the evaporation
process is slowed down resulting in smaller loss of mass and hence forcing 
it to fall on the ground under the action of gravitation. 
Comparison of results for droplet of initial radius $50\mu$m
for $\textrm{RH}=40\%, 60\%$ and $80\%$ 
indicate that the effects of $\textrm{RH}$ is significant. 
The droplets with initial radius 25 $\mu$m or less, however, 
remain suspended in the air, namely, these droplets  
do not settle under the action of gravitation but continue to diffuse in the air making
the use of mask mandatory to prevent the infection. 
It needs to be reiterated at this point that we follow the 
trajectories of droplets of radius upto $2 \mu${\text m} 
created by the process of evaporation.
At higher temperature  the evaporation becomes faster as can 
be seen from the results displayed in Fig~\ref{fg2}.
A droplet of initial radius $50 \mu$m evaporates swiftly at $35^\circ$ C than
at $20^\circ$ C at the same relative humidity of $40\%$.
It is observed that at higher $\textrm{RH}$ and low $T$ the effect of evaporation is small and
hence the droplets survive for longer time. 
The relation between the climatic condition and 
index of airborne infection rate and concentration rate 
of particle in saliva can be found in Ref.\cite{dbouk3} 
(see also \cite{dbouk2}).  

\begin{figure}[ht]
\includegraphics[width=7.0cm]{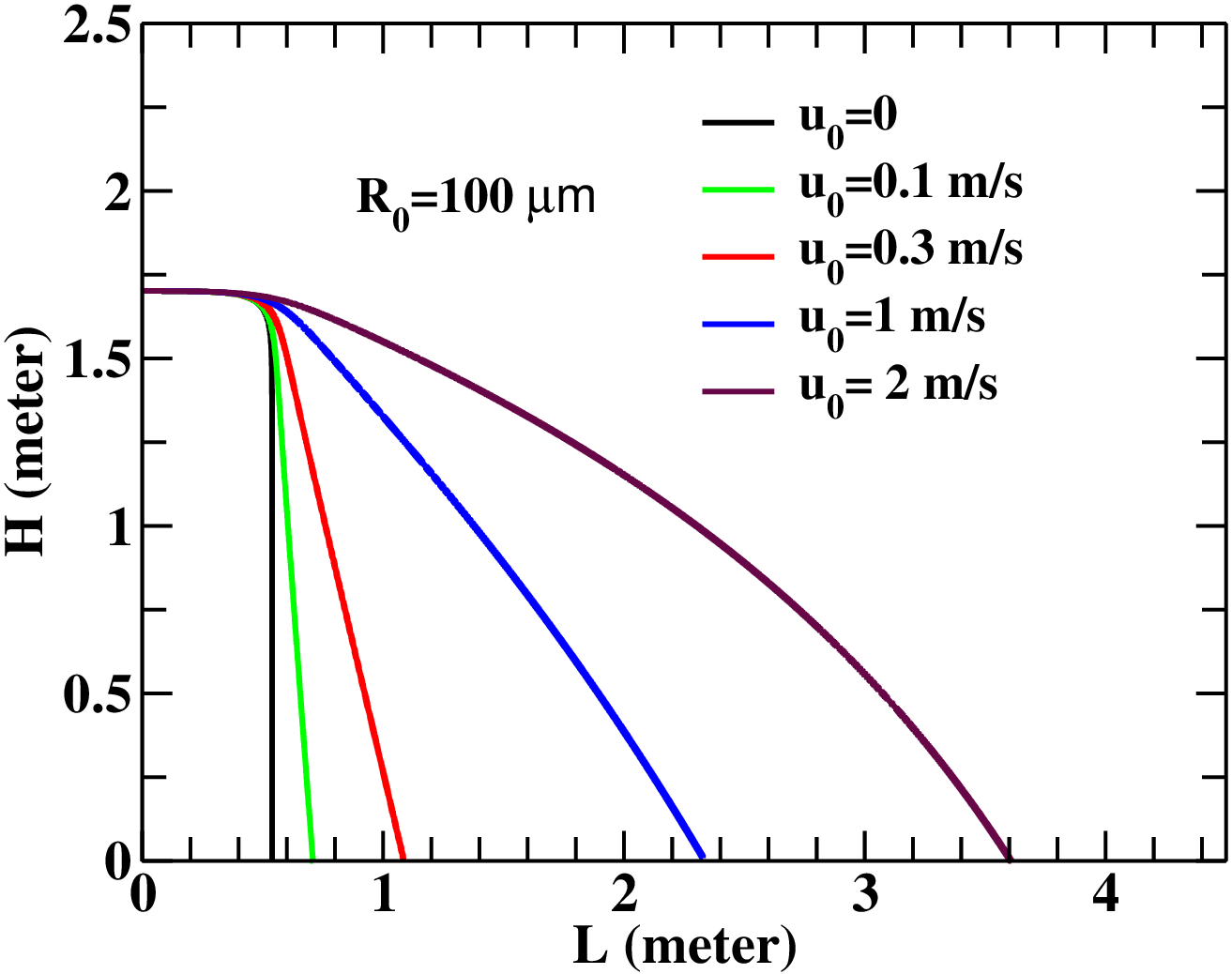}
\includegraphics[width=7.0cm]{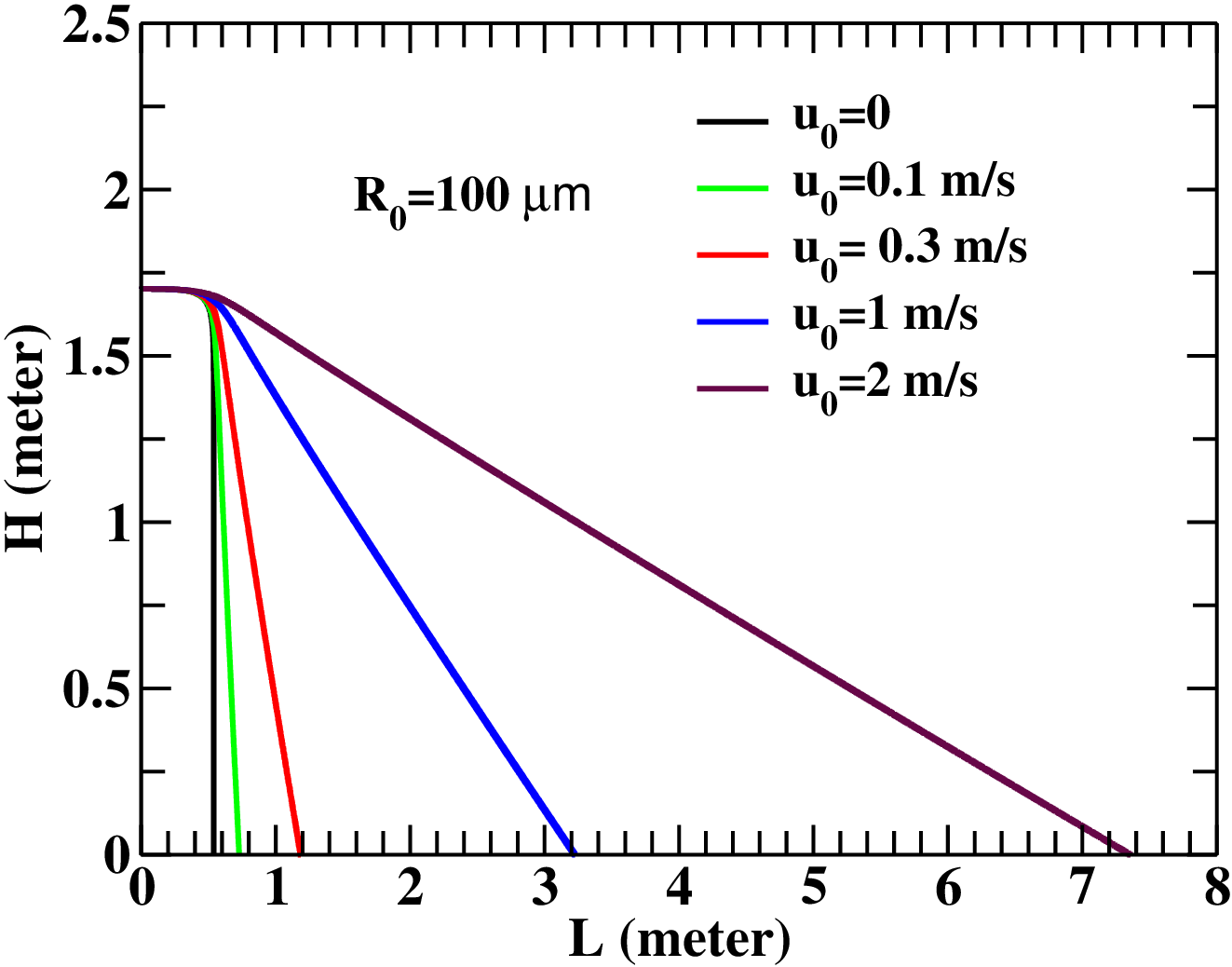}
\caption{(Color online) 
Left panel:The change in the height, $H$, as a function of $L$ 
for a droplet of radius 100 $\mu$m for different peak values ($u_0$)
of the wind velocity profile at $T=20^\circ$ and $\textrm{RH}=60\%$. 
The droplet is ejected in a room of size 5 meter 
with initial ejection velocity is 21 m/s. 
Right panel: Same as left panel for different wind velocity 
of constant values (independent of space coordinate) as indicated in the figure. 
}
\label{fg2a}
\end{figure}
The effects of wind flow in indoor (left panel)
and outdoor (right panel) conditions have been 
demonstrated in Fig.~\ref{fg2a}. The left panel shows
the results for flow profile mentioned 
above for various values of $u_0$ in 
a room of size 5 metre (which sets the value
of $x_{\text{max}}$). The right panel 
indicates the results for different wind flow with
constant speed (independent of space coordinate) in an open air, say.
It is clearly observed that the wind flow substantially 
affects the  distance traveled by the droplets. We find that 
a droplet of radius 100 $\mu$m ejected with a velocity 21 m/s 
traverses a distance of 3.6 meter
(for velocity profile mentioned above with $u_0=2$ m/s) and as large as 7.4
meter for constant velocity of magnitude 2 m/s at $T=20^\circ$C and 
$\textrm{RH}=60\%$.  
This indicates that
the droplet released by an infected individual can travel distances
much larger than prescribed 2 meters if the wind velocity is large. 
\begin{figure}[ht]
\includegraphics[width=18pc,clip=true]{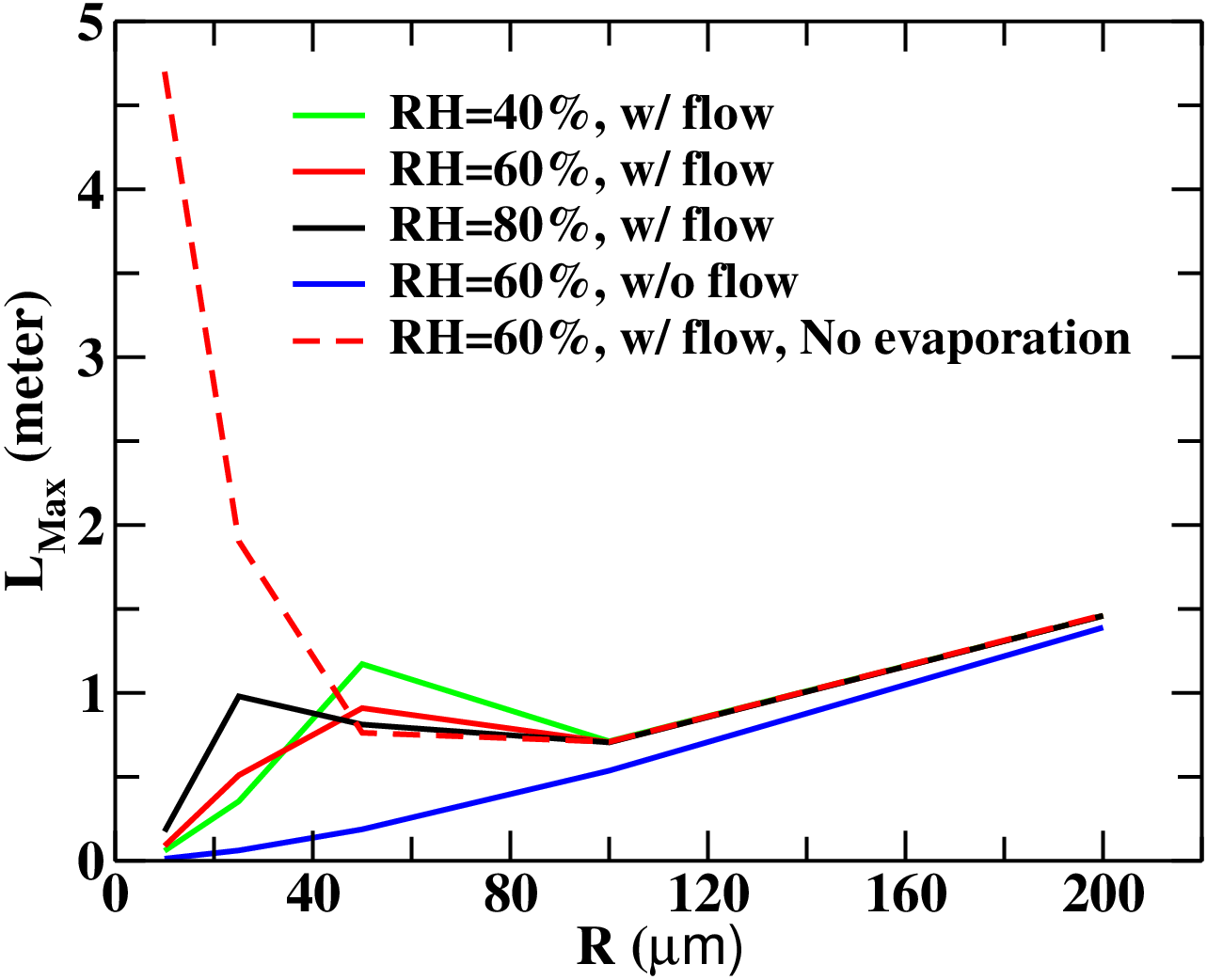}\hspace{2pc}
\caption{The variation of the maximum horizontal distance ($L_{max}$) traveled by
the droplets as a function of the droplet radius for 
different relative humidity have been shown. 
Here initial ejection velocity, $V_0=21$ m/s and the 
peak value of the wind flow velocity, $u_0$=0.1 m/s.}
\label{fg3}
\end{figure}

In Fig~\ref{fg3}, we have displayed the  variation of the maximum horizontal distance traveled by
droplets as a function of radius at different $\textrm{RH}$. 
The droplets of  small and intermediate sizes are 
strongly influenced by the $\textrm{RH}$. We notice mild effects of flow (for small
$u_0$) and $\textrm{RH}$ on bigger droplets. 
However, the droplets with intermediate radii are influenced  the most due to 
variation in $\textrm{RH}$ and flow mainly due to their longer life time 
(from ejection to the formation of aerosol of size $2 \mu$m) in the air than 
the smaller droplets which are subjected to quick evaporation.

\begin{figure}[ht]
\includegraphics[width=18pc,clip=true]{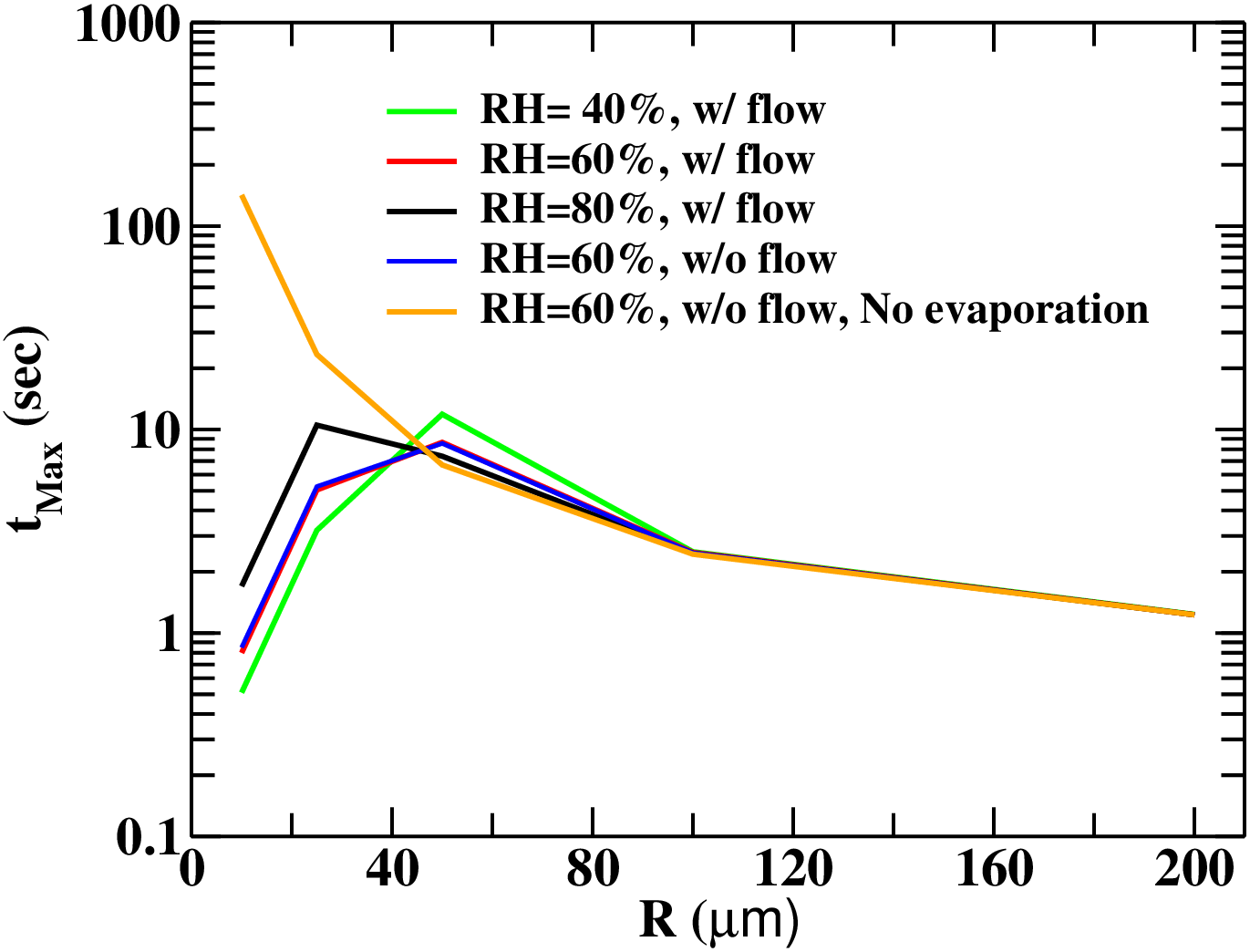}\hspace{2pc}
\caption{The variation of the maximum time the droplets take to settle on the ground 
under the action of gravity and the time the smaller droplets  take to 
evaporate to generate aerosols of average radii $2\mu$m are displayed here.
The sensitivity of the results on the relative humidity has also been shown
for $u_0=0.1$ m/s.}
\label{fg4}
\end{figure}

\begin{figure}[ht]
\includegraphics[width=18pc,clip=true]{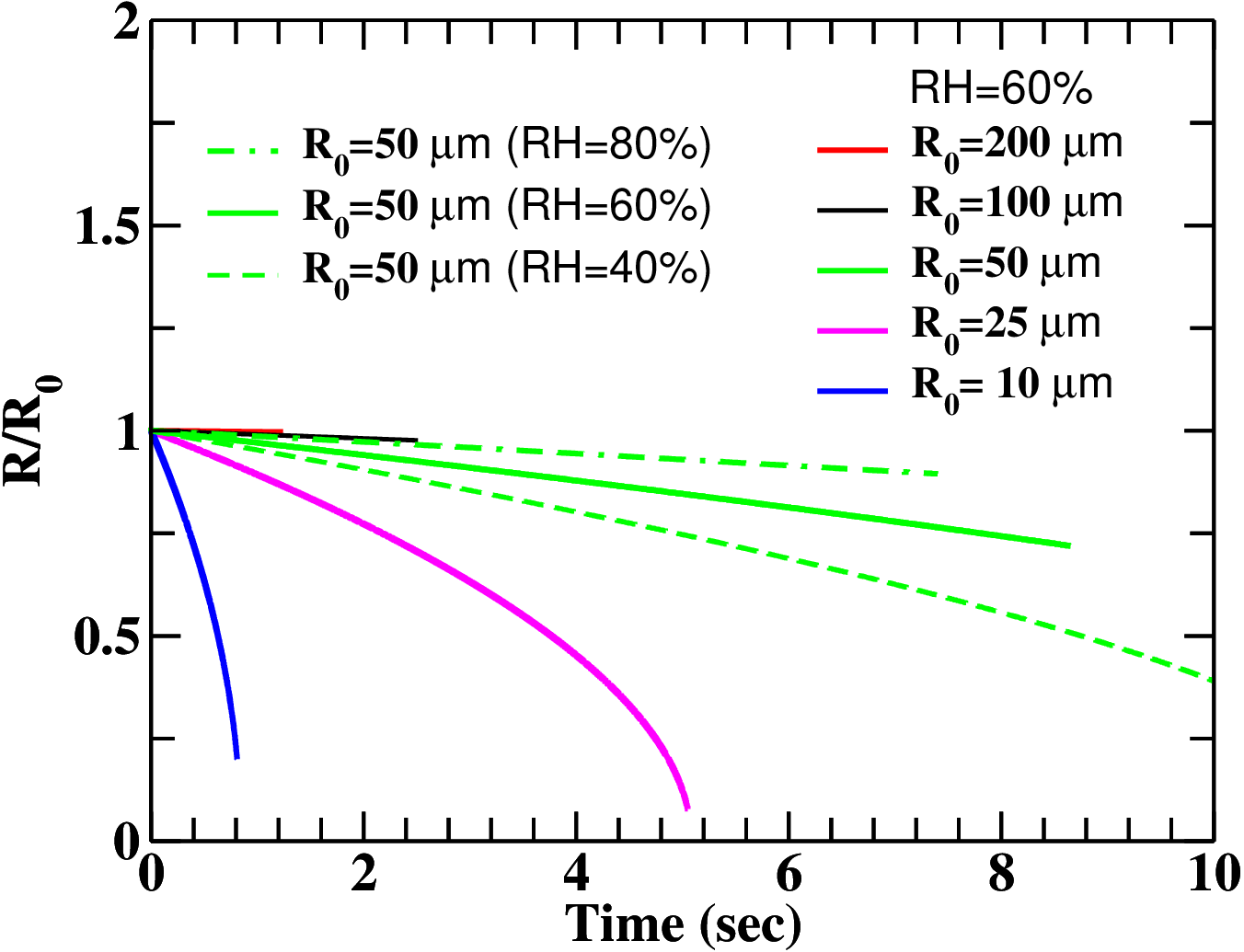}\hspace{2pc}
\caption{Change in the running radius, $R(t)$ normalized to the initial radius
have been depicted as a function of time for different relative humidity.
Here $R_0$ is the initial radius of the droplets.
The value of  $V_0$ and $u_0$ are taken as $21$ m/s and $0.1$ m/s respectively.}
\label{fg5}
\end{figure}

\begin{figure}[ht]
\includegraphics[width=18pc,clip=true]{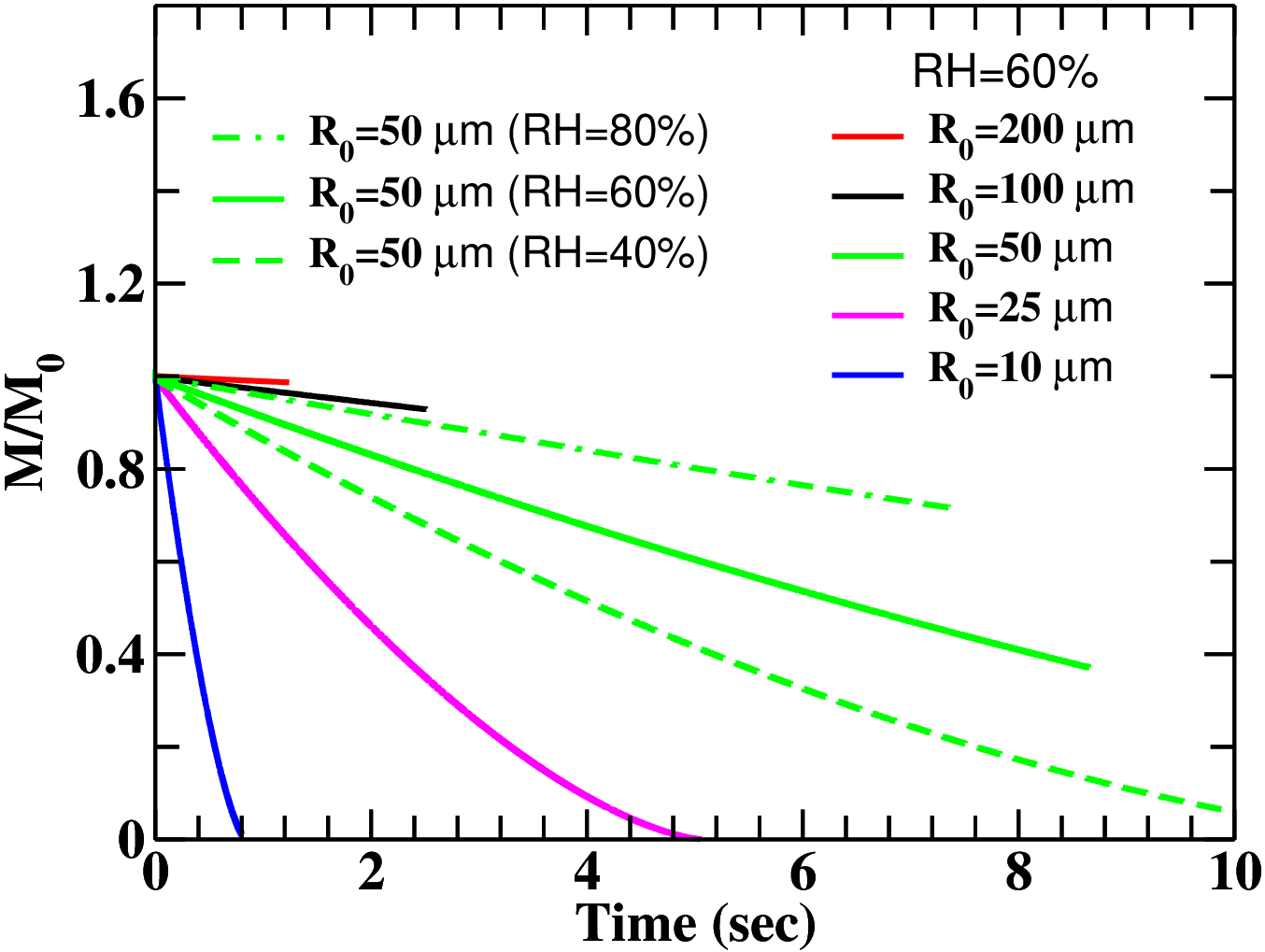}\hspace{2pc}
\caption{Same as Fig.~\ref{fg5} for running mass to initial mass of the
droplets.
}
\label{fg6}
\end{figure}

Now we would like to estimate the  maximum time of suspension of the evaporating droplets 
of different radii in the air before  settling gravitationally 
at different weather conditions.
Relevant results  to address this issue are displayed in Fig.~\ref{fg4}. 
The effects of evaporation is found to be significant for small droplets. 
At higher $\textrm{RH}$ the droplets
have the higher survival probability due lesser mass loss
by evaporation and hence they travel larger distance in the
air.   However, the time of suspension 
of intermediate and smaller sized droplets are very sensitive 
to the evaporation.   
Through the process of evaporation smaller droplets may generate isolated virus 
which may survive for more than an hour~\cite{swxong}.
However, droplets having larger radius fall on the ground quickly  
(smaller $t_{\text{max}}$) 
due to stronger gravitational force compare to the diffusive force. 

The sensitivity of evaporation on $\textrm{RH}$ is clearly demonstrated 
through results depicted in Figs.~\ref{fg5} and \ref{fg6}. 
In Fig~\ref{fg5}, we have plotted the change in the running radius $(R)$ of
the droplets normalized to their initial values ($R_0$) for different  humidity.
The reduction of mass due to evaporation 
is lower at higher $\textrm{RH}$ which allows the droplet to survive longer. 
It is noticed that smaller droplets are very sensitive to 
evaporation and they  generate droplet nuclei before reaching the ground.
In Fig~\ref{fg6}, the change in the running masses ($M$) of the droplet normalized to their 
initial masses ($M_0$) is illustrated as a function of time for 
different humidity. The results are consistent with those displayed in Fig.~\ref{fg5}.
The results clearly indicate the effects of $\textrm{RH}$ on the evolution of the 
droplets, for example a droplet of radius $50 \mu$m, will 
approximately loss $60\%$ and $20\%$ of its mass due to evaporation at 
$\textrm{RH}=60\%$ and $80\%$  respectively.

\section{Summary and Discussions}
We have investigated the evolution of the droplets 
ejected during coughing, sneezing or speaking by solving
the Langevin stochastic differential equation with the inclusion of evaporation  
process. The drag, diffusive  and the gravitational forces 
are included in this study. The correction to the drag force 
in the non-laminar (turbulent) flow region 
implied by large Reynolds number has been 
implemented here by using the parameterization of the 
Reynold's number dependence of the  drag coefficients. 
The droplets of various sizes have been considered with large 
ejection velocity (21 m/s) to get the  upper limit of the distance 
travel by the droplets. The effects of different weather conditions have
been taken into consideration through temperature, relative humidity and wind
flow. It is found that the maximum distance that a large droplet
of size 200 $\mu$m travels is about 1.5 meter ($u_0=0.1$ m/s) which is
not affected significantly by low wind velocity. The tinier droplets 
diffuse  through the air for long duration~\cite{Bouroaiba} due to weaker
gravitational influence.  

An infected person emits droplets of varying sizes. 
In addition to direct emission, smaller droplets 
can  be created by the mechanism of evaporation
(see also~\cite{FanLia}) and 
fragmentation ~\cite{xuwang}. 
It is found that droplets of smaller size 
evaporate to create aerosols, smaller the sizes 
quicker they evaporate. 
Isolated virus generated by the process of evaporation 
can survive in the air for more than an hour~\cite{swxong}. 
The evolution of these droplets will be
controlled by the diffusive forces and such droplets may remain 
suspended in the air for hours. 
However, it has also been reported~\cite{jama} that a multi-phase 
turbulent gas cloud 
is produced along with the droplets at the time of coughing and sneezing.
The droplets within the envelope of this gas cloud may evade evaporation
and prolong their survivability as isolated droplets. Droplets
of larger size (say, 100 $\mu$m) can travel a distance
as large as 3.6 meter in indoor condition (for the 
flow profile defined above) and even more
in outdoor condition ({\it i.e.} constant flow velocity) if the wind flow is strong.
It is appropriate to mention at this juncture that 
the intermediate size droplets are very 
sensitive to the weather conditions.  The climatic 
conditions  determines whether the intermediate 
size droplets will settle in the ground 
under the action of gravity or evaporate to aerosol. 
This indicates that the weather condition play crucial role in deciding  
the severity of spreading of the infection.
Therefore, the  mask and face shield
should be used to prevent the virus ~\cite{Howard,pnas,verma}. 
The maintenance of ventilation in indoor situation is also
very crucial to avoid the infection~\cite{NSen,SBathula}. 

The maintenance of
only a social distance of 2 meter as norm to avoid the virus is
not corroborated by  the present investigation. Because the aerosol
created by the evaporation may remained suspended in the air for
long duration and under strong wind flow the droplets can travel
distance much larger than 2 meter.  The trajectories
of the small droplets/aerosols are determined by the diffusive force as
it dominates over the gravitational force. These trajectories
are highly zig-zag in nature and therefore, although the path length
traversed  
may be large but their mean values  are small and
such droplets remain suspended for long time. 
The calculations based on Wells-Riley probability  
and computational fluid dynamics~\cite{AFoster} suggest 
that the use of mask in class room environment 
will be more  effective than maintaining
the physical distancing to escape the virus. 
In two Wuhan hospitals the micrometer and sub-micrometer droplets
of Sars-COV-2 were found at a distance of about 3 meters 
away from the bed of an infected person~\cite{Liu}.


\section*{Data Availability Statement}
The data that supports the findings of this study are available within the article.

\section*{Declaration of competing interest}
The authors declare that they have no known competing financial interests or personal 
relationships that could have appeared to influence the work reported in this paper.

\section*{Authors contributions} All the authors contributed equally. 

\section*{Acknowledgement}  
SKD would like to acknowledge IIT Goa for internal funding (No. 2020/IP/SKD/005)
and Professor Barada Kanta Mishra for useful discussions and encouragement. 
We thankfully acknowledge fruitful discussions with Raja Mitra and Ashish Bhateja.

\end{document}